# Enabling Complex Wikipedia Queries - Technical Report

## Gilad Katz and Bracha Shapira

### Department of Information Systems Engineering,
### Ben-Gurion University of the Negev, Israel

**Abstract**

In this technical report we present a database schema used to store Wikipedia so it can be easily used in query-intensive applications. In addition to storing the information in a way that makes it highly accessible, our schema enables users to easily formulate complex queries using information such as the anchor-text of links and their location in the page, the titles and number of redirect pages for each page and the paragraph structure of entity pages. We have successfully used the schema in domains such as recommender systems, information retrieval and sentiment analysis. In order to assist other researchers, we now make the schema and its content available online.

## 1. Introduction

Wikipedia, the online encyclopedia, has become one of the most popular websites in the world.[1] Millions of people rely on it every day for the discovery and validation of information. In addition to becoming a source of information for the general population, Wikipedia has also become an important research tool, as well as a subject of research.

The statistics on Wikipedia's size and scope are truly impressive: there are currently 287 different versions of Wikipedia, each in a different language. The English Wikipedia, by far the largest, currently contains over 4.6 million pages (see Figure 1). Although one might expect the rate of the addition of new entries to have slowed over time, it has remained relatively steady at approximately 30,000 additions per month since 2011 (see Figure 2).

Wikipedia has been used in multiple scientific fields: computer science, medicine, physics, sociology etc. According to [1], more and more of Wikipedia-related papers seem to be generated with each passing year. Wikipedia has several traits which constitute it as such a valuable source of information for research:

*Size and scope* - As mentioned above, the English Wikipedia alone has over 4.6 million entries. Encyclopedia Britannica, one of the best-known "regular" encyclopedias, has 40,000. This great difference in scope suggests that Wikipedia covers a multitude of fields and areas of interest that are not covered by curated encyclopedias.

*Timely and updated* – Because of Wikipedia's open editing policy which enables any person to modify its content, the information it contains is almost always up-to-date. Case in point: In 2013, a few minutes after the election of the new pope, one of the authors of this study reviewed the relevant Wikipedia entries and found them to already be updated with the elected pope's new status.

---

[1] http://www.alexa.com/topsites

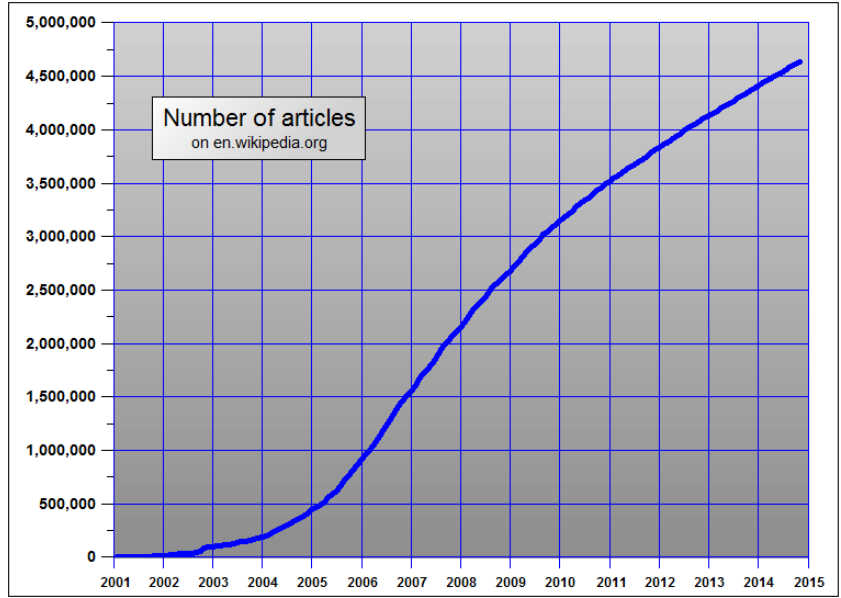
Fig. 1. The number of articles in the English Wikipedia over the years (taken from Wikipedia's website[2])

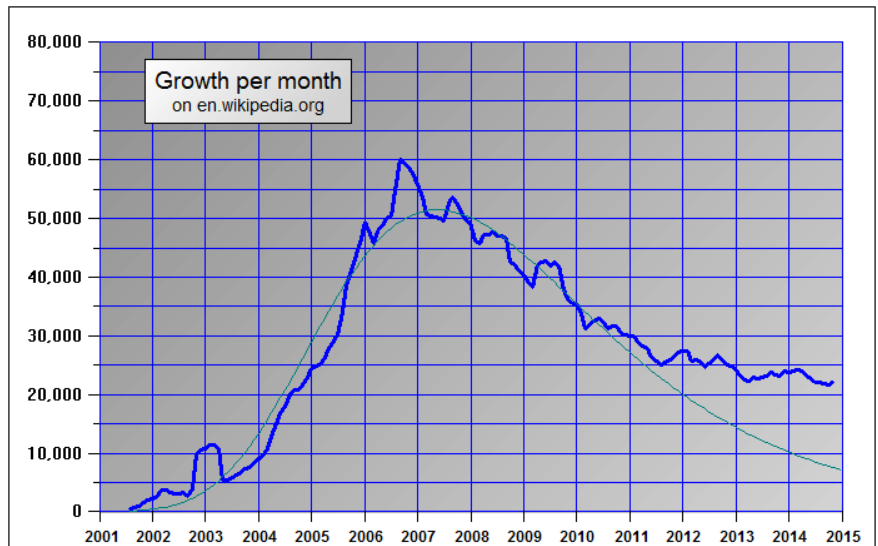
Fig. 2. The number of new articles added to the English Wikipedia each month (taken from Wikipedia's website)

*Tags and meta-data* – Wikipedia contains multiple types of user-generated content (UGC); categories, links, redirect pages and infoboxes can all be used to infer the type, attributes and connections among the various entities represented in Wikipedia.

*Wisdom of the crowd* – Since every person has the ability to contribute content to Wikipedia, it reflects the thoughts, ideas and perceptions of peoples, groups and societies [2]. This enables us to use Wikipedia to measure popularity, importance and influence. In a sense, Wikipedia is "representative of the real world".

In this technical report we present a schema designed to facilitate the utilization of Wikipedia information. The said schema provides easy access to the most frequently used types of Wikipedia information: page titles, links, categories, redirect pages and paragraphs structure. As we demonstrate later on, the use of simple SQL queries can provide users with

---
[2] http://en.wikipedia.org/wiki/Wikipedia:Size_of_Wikipedia

information that previously was far harder to obtain. In order to assist other researchers who wish to integrate information from Wikipedia in their work, the schema and its content can be downloaded in the following link (if the schema is used in any academic work, please cite this technical report): http://www.ise.bgu.ac.il/ECWQ/

**2. The Wikipedia Schema**

We chose to represent the information extracted from Wikipedia using a relational database. The main reason for this was the fact that this enabled us to take advantage of all the benefits offered by such databases – primary and foreign keys, the verification of the data that was inserted into the database and the utilization of indices for faster query execution. Moreover, despite being relatively large (~50GB, with about half of the size taken up by the indices that accelerate large extractions), a database of this size can be conveniently used on a stand-alone server (or even a laptop) without resorting to big-data or no-sql-based solutions. As Wikipedia's growth is linear rather than exponential, we currently do not see the need to use solutions such as Hive.

The schema is presented in Figure 3. We now go over the tables and describe their fields:
1) **Tbl_Wiki_Page** – this is the main table of the schema. It contains information on every Wikipedia page, including categories and redirect pages. It Contains the following fields:
    - **Page_id** – the internal id of the page. This field is unique.
    - **Page_name** – the text that appears at the title of the page
    - **File_path** – the path to the location of a *.txt* file that contains the text of the page (in addition to the schema, we also provide a directory containing the texts of all Wikipedia pages).
    - **Page_type** – this field has three possible values: 1 (entity pages), 2 (redirect pages) and 3 (category pages)
    - **Stemmed_name** – the title of the pages after being stemmed using the Lovins Stemmer [3] and stop-words removal. This field is important for information retrieval tasks, as it enables easy matching to the analyzed text, which is also often stemmed.
2) **Tbl_Wiki_Page_Redirect** – this table provides information of the *redirect pages* (pages that upon accessing them the user is immediately transferred to another page). It contains the following fields:
    - Page_id – the unique identifier of the redirect page
    - Redirected_page_title – the title of the page the redirect page is *pointing* to
    - Redirected_page_id – the id of the *entity page* the redirect page is pointing to
3) **Tbl_Wiki_Page_Categories** – for each page, this table contains all its categories. It is important to note that each page will appear multiple times. For example, a page with five categories will appear five times. It contains the following fields:
    - Page_id – the id of the page to which we assign the category
    - Category_page_id – the id of the category page
4) **Tbl_Wiki_Page_Links** – this table contains information about the links found in each Wikipedia page. Each link in Wikipedia has a separate tuple in this table. It contains the following links:
    - Page_id – the ID of the page containing the link
    - Link_page_id – the ID of the page to which the link redirects
    - Pos_in_page – the character offset of the link. This is counted from the top of the page until the first word of the link's *anchor text*
    - Link_description – this is the link's anchor text – the text that is highlighted as a link

- Stemmed_link_description – the link's anchor text after undergoing stemming using the Lovins stemmer. As in the *stemmed_name* field in *Tbl_Wiki_Page*, this field was created in order to simplify the use of this schema in information retrieval tasks

5) **Tbl_Wiki_Page_Paragraphs** – this table contains information on the paragraph structure and hierarchy of the Wikipedia page. Our rationale for creating this table was the assumption that content that appears in the first paragraphs is more important that appears in the last (particularly in detailed Wikipedia entries). Similarly, we assumed that a piece of information that appears in a sub-paragraph is less important than one that appears at the top paragraph of the hierarchy. This table enables us to model these relationships. It contains the following fields:
    - Page_id – the unique identifier of the Wikipedia page
    - Paragraph_id – a running index of all the paragraphs in the page
    - Paragraph_start_pos – the *characters offset* of the first letter in the paragraph from the top of the page
    - Paragraph_end_pos – the character offset of the last letter in the paragraph from the top of the page
    - Paragraph_level – the depth in the hierarchy of the paragraph. Paragraph without a "parent" have the value of 1.

The link for downloading the schema is located at the end of section 1.

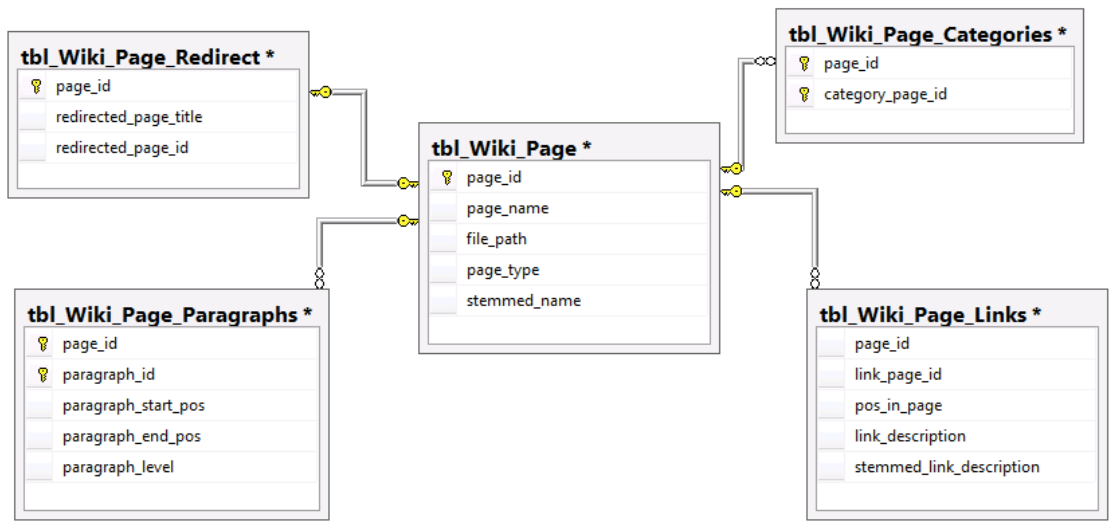

Fig. 3. The Wiki-DB schema

3. **Use-Case Queries**

In this section we present several queries that were utilized in our studies over the past few years [4-7], as well as a those that are used in current work. For each example we first state the question we wish to answer and then present the SQL query that provides the answer.

a) Obtain the ID and title of Wikipedia pages that share a specific category
   select page_id, page_name from tbl_Wiki_Page where
        page_id in (select page_id from tbl_Wiki_Page_Categories where category_page_id = 691014)

b) Given two groups of Wikipedia pages, find all the pages in group1 that contain links to pages in group2 (group1 and group2 are in tables tbl_Temp1 and tbl_Temp2, respectively)
```sql
select distinct page_id, link_page_id from tbl_Wiki_Page_Links
    where page_id in (select field1 from tbl_Temp)  and link_page_id in
    (select field1 from tbl_Temp2)
```

c) Obtain the number of redirect pages for a set of Wikipedia pages
```sql
select redirected_page_id, COUNT(*) from tbl_Wiki_Page_Redirect
    where page_id in (select field1 from tbl_Temp)
    group by redirected_page_id
```

d) Obtain the IDs of all Wikipedia pages that the a specific page points to in its *second* paragraph
```sql
select link_page_id from tbl_Wiki_Page_Links
    where pos_in_page >= (select paragraph_start_pos from
    tbl_Wiki_Page_Paragraphs where page_id = 12 and paragraph_id = 1)
    and pos_in_page <= (select paragraph_end_pos from
    tbl_Wiki_Page_Paragraphs where page_id = 12 and paragraph_id = 1)
    and page_id = 12
```

While these are only a few examples, they clearly illustrate the effectiveness and ease of use of the presented schema.

## 4. Conclusion

In this technical report we present a database schema for Wikipedia which goes a long way in making Wikipedia more accessible to researchers. As we demonstrate in the previous section of this report, our schema makes it far easier to perform complex queries and obtain information that previously would have required a great deal of time and effort on one's behalf.

Despite the advantages of the proposed schema, we must also point out its major shortcoming – the fact that it is static. We are currently working on improving and documenting the code that processes the Wikipedia dump file so that researchers are able to create up-to-date versions of our schema.